\documentclass[aps,prd,showpacs,notitlepage,nofootinbib,superscriptaddress,floatfix,showkeys,twocolumn]{revtex4-1}

\usepackage{blindtext}
\usepackage{hyperref}
\usepackage{amsmath,amssymb}
\usepackage{float}
\usepackage{microtype}
\usepackage{ragged2e}  

\usepackage{graphicx}
\usepackage{bm}
\usepackage{latexsym}
\usepackage{epsfig}
\usepackage{psfrag}
\usepackage[dvipsnames]{xcolor}
\usepackage{subfigure}
\usepackage{subcaption}
\usepackage{graphicx}
\usepackage{multirow}
\usepackage{amssymb}
\usepackage{amsmath}
\usepackage{bm}
\usepackage{latexsym}
\usepackage{epsfig}
\usepackage{psfrag}
\usepackage[normalem]{ulem}
\usepackage{textcomp}
\usepackage{color}
\usepackage{pstricks}
\usepackage[utf8]{inputenc}
\usepackage{comment}
\usepackage{hyperref}
\usepackage[mathlines]{lineno}
\usepackage{hyperref}
\usepackage[all]{hypcap}
\hypersetup{  
    colorlinks=true,
    linkcolor=blue,
    filecolor=red,      
    urlcolor=cyan,
    citecolor=green,
    }

\def\beq{\begin{equation}}
\def\eeq{\end{equation}}
\def\bea{\begin{eqnarray}}
\def\eea{\end{eqnarray}}
\def\be{\begin{equation}}
\def\ee{\end{equation}}

\def\bse{\begin{subequations}}
\def\ese{\end{subequations}}

\def\ee{\eta_{\rm e}}

\def\Mpl{M_{_\mathrm{P}}}

\def\d{\mathrm{d}}

\def\w{w_{\phi}}

\begin{document}

\title{Minimal Plateau Inflation in light of ACT DR6 Observations}
\author{Md Riajul Haque}
\email{E-mail:riaj.0009@gmail.com }
\affiliation{\,Physics and Applied Mathematics Unit, Indian Statistical Institute, 203 B.T. Road, Kolkata 700108, India}
\author{Debaprasad Maity}
\email{E-mail:debu@iitg.ac.in }
\affiliation{\,Department of Physics, Indian Institute of Technology Guwahati, Guwahati, Assam, India}

\begin{abstract}
We explore a class of minimal plateau inflationary models constrained by the latest cosmological observations from ACT DR6, Planck 2018, BICEP/Keck 2018, and DESI, collectively referred to as P-ACT-LB-BK18. These models, characterized by a non-polynomial potential, are analyzed using both inflationary and post-inflationary reheating dynamics, and the limits on the viable model parameter space are obtained. Our results show that the minimal model 
with matter-like post inflationary reheating phase remains consistent with current data at both $1\sigma$ and $2\sigma$ levels. The inflaton potential's exponent $n$ and reheating epoch are intertwined in that upon its increase, corresponding to the stiffer reheating equation of state, the viable model parameter space in accordance with ACT shrinks, which is further facilitated by the primordial gravitational waves (PGWs) overproduction. We further explored a supergravity-inspired extension of the model under study with similar results, but with tighter constraints on the model parameters. These results emphasize the importance of jointly analyzing CMB data and reheating physics to test inflationary models.
\end{abstract}

\maketitle

\black

\section{Introduction}
\label{sec:intro}
Inflation stands as the most successful paradigm for describing the dynamics of the early universe and the observed large-scale structure. Observations of the Cosmic Microwave Background (CMB) \cite{ACT:2025fju,ACT:2025tim} provide some of the most stringent constraints on the nature of the inflaton field responsible for driving this accelerated expansion. Complementary to this, measurements of baryon acoustic oscillations (BAO) \cite{DESI:2025zgx} in the large-scale distribution of matter have further reinforced inflation as a compelling framework for early universe cosmology. \\
Over the years, a wide variety of inflationary models have been proposed, including chaotic~\cite{Linde:1983gd,Martin:2013tda}, hybrid~\cite{Linde:1993cn}, Higgs, Starobinsky ~\cite{Starobinsky:1980te,Kallosh:2013hoa,Ellis:2013nxa,Kallosh:2013yoa}, axion~\cite{Freese:1990rb}, and $\alpha$-attractor models~\cite{Kallosh:2013yoa,Kallosh:2013hoa,Roest:2013fha,Ferrara:2013rsa,Kallosh:2013tua,Cecotti:2014ipa,Kallosh:2014rga,Galante:2014ifa}, among others. Despite their diversity, many of these models, under the slow-roll approximation, make predictions that lie within a narrow range of observable parameters—most notably, the scalar spectral index $n_s$ and the tensor-to-scalar ratio $r$. As a result, increasingly precise measurements of these quantities have become crucial for distinguishing between competing inflationary scenarios.\\
Starting with COBE \cite{Bennett:1996ce,Hauser:1998ri}, observations of the cosmic microwave background (CMB) have progressively improved in precision over the past three decades, culminating in the latest results from the Atacama Cosmology Telescope (ACT) \cite{ACT:2025fju,ACT:2025tim}. The latest Data Release 6 (DR6) from the Atacama Cosmology Telescope (ACT) has substantially improved the precision of high-$\ell$ CMB power spectrum measurements. When combined with Planck data (P-ACT), these results yield a refined estimate of the scalar spectral index: $n_s = 0.9709 \pm 0.0038$ \cite{ACT:2025fju,ACT:2025tim}. The inclusion of CMB lensing and Baryon Acoustic Oscillation (BAO) data from DESI (P-ACT-LB) further sharpens this constraint to $n_s = 0.9743 \pm 0.0034$ \cite{ACT:2025fju,ACT:2025tim}, marking a $2\sigma$ shift relative to Planck-only measurements.\\
Motivated by these developments, a wide array of inflationary models have been proposed, while many existing frameworks have undergone renewed scrutiny and refinement \cite{Kallosh:2025rni,Aoki:2025wld,Berera:2025vsu,Dioguardi:2025vci,Gialamas:2025kef,Salvio:2025izr,Antoniadis:2025pfa,Kim:2025dyi,Dioguardi:2025mpp,Gao:2025onc,He:2025bli,Pallis:2025epn,Drees:2025ngb,Haque:2025uis,Haque:2025uri,Yin:2025rrs,Byrnes:2025kit,Maity:2025czp,Mondal:2025kur,Peng:2025bws,Yi:2025dms,Gialamas:2025ofz,Yogesh:2025wak,Kallosh:2025ijd}. One compelling approach involves coupling the inflaton linearly to curvature through a term such as $\xi \phi R$. For $\xi \sim \mathcal{O}(1)$, such models have been shown to fit ACT data well \cite{Kallosh:2025rni}. However, these non-minimal coupling models can be reformulated as minimal ones via appropriate conformal transformations. Motivated by this, we reinvestigate a new class of minimal inflationary models that yield a plateau-like potential at large field values~\cite{Maity:2019ltu,Maity:2016zeu}. By jointly analyzing inflationary dynamics and the post-inflationary reheating phase, we find that scenarios with a matter-like reheating equation of state ($w_\phi = 0$) are most consistent with current observational data.

This paper is organized as follows: In Section \ref{sec:inflation_overview}, we introduce the minimal and supergravity-inspired inflaton potentials and derive the corresponding inflationary observables. Section \ref{sec:reheating_GW} details the dynamics of the post-inflationary reheating phase and examines the constraints arising from the overproduction of primordial gravitational waves (PGWs), especially through bounds on $\Delta N_{\rm eff}$. In Section \ref{sec:results}, we present a comprehensive numerical analysis of the allowed parameter space using the most recent P-ACT-LB-BK18 dataset. Finally, we summarize our main findings and outline future directions in Section \ref{sec:conclusion}. 
\section{Minimal inflation models}
\label{sec:inflation_overview}
In this section we introduce two sets of models following \cite{Maity:2019ltu,Maity:2016zeu}. In the first set of models, we consider the following phenomenological form of the non-polynomial inflaton potential, 
\begin{equation} \label{model1}
V(\phi) =  \frac{\Lambda \phi^n}{1+\left(\frac{\phi}{\phi_*}\right)^n} . 
\end{equation}
The parameter $\Lambda$ sets the scale of inflation. The index $n$ is assumed to be an even integer \cite{Maity:2016zeu}. A new mass scale $\phi_{\ast}$ is introduced that controls the shape of the potential. For large field $\phi >>\phi_*$ the potential becomes flat with a constant potential value $\Lambda \phi_*^n$. 

\begin{figure*}
\includegraphics[width=0017.50cm,height=013.0cm]{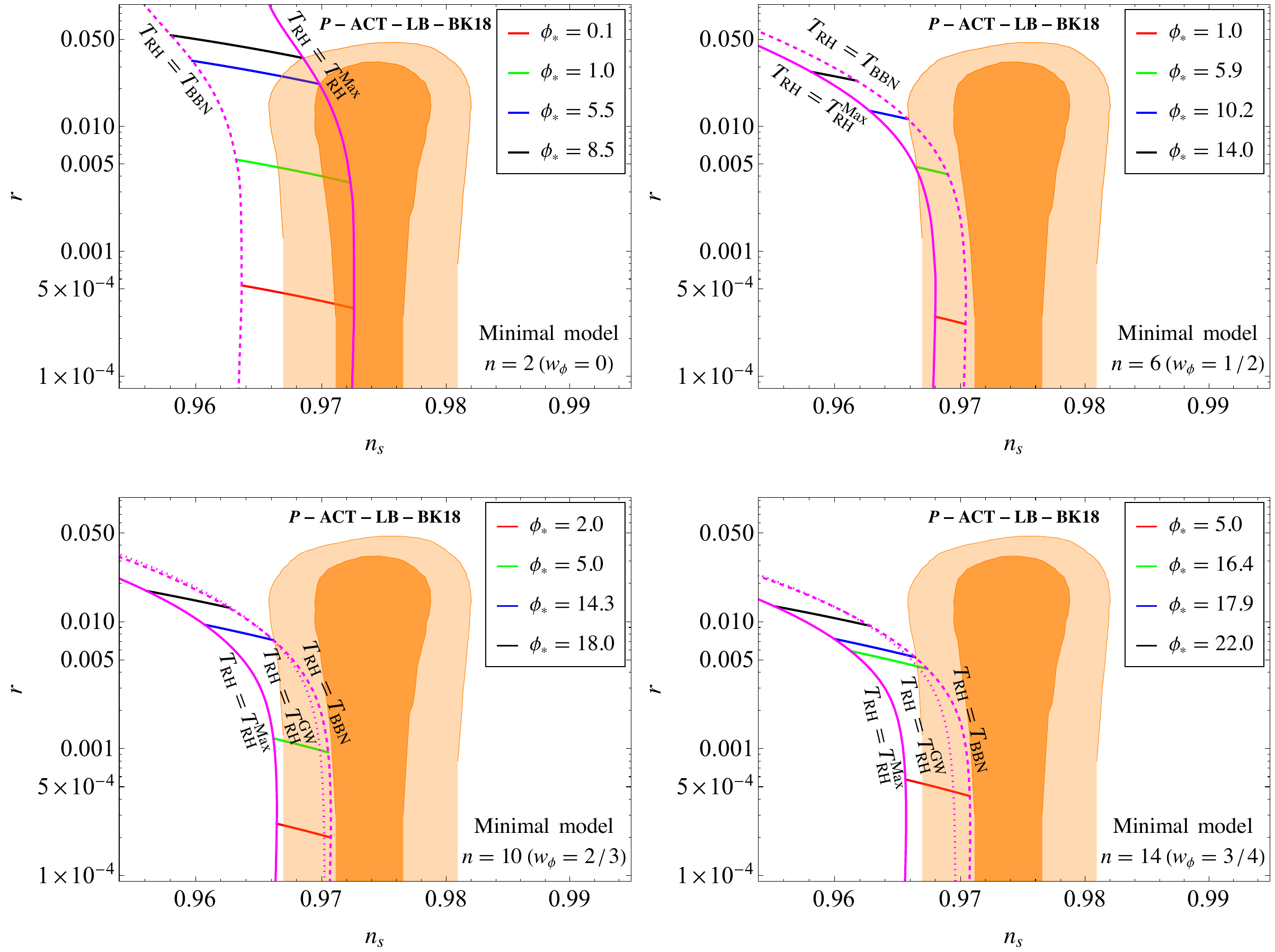}
\caption[]{\justifying \em  Predictions from the minimal model across different values of $(\phi_,, w_\phi)$ in $(n_s-r)$ plane, alongside the most recent combined observational bounds from \textsc{P-ACT-LB-BK18}. Shaded regions in dark and light orange represent the 68\% ($1\sigma$) and 95\% ($2\sigma$) confidence levels, respectively. The reheating temperature varies within the range $T_{\rm BBN} \leq T_{\rm RH} \leq T_{\rm RH}^{\rm Max}$, denoted by dashed and solid magenta lines. A dotted magenta line indicates the critical temperature $T_{\rm RH}^{\rm GW}$, inferred from constraints on $\Delta N_{\rm eff}$ to prevent excessive production of PGWs. The parameter $\phi_*$ is given in units of the reduced Planck mass, $M_{\rm p}$.}
\label{ns-r:minimal}
\end{figure*}

\begin{figure*}
\includegraphics[width=0017.50cm,height=013.0cm]{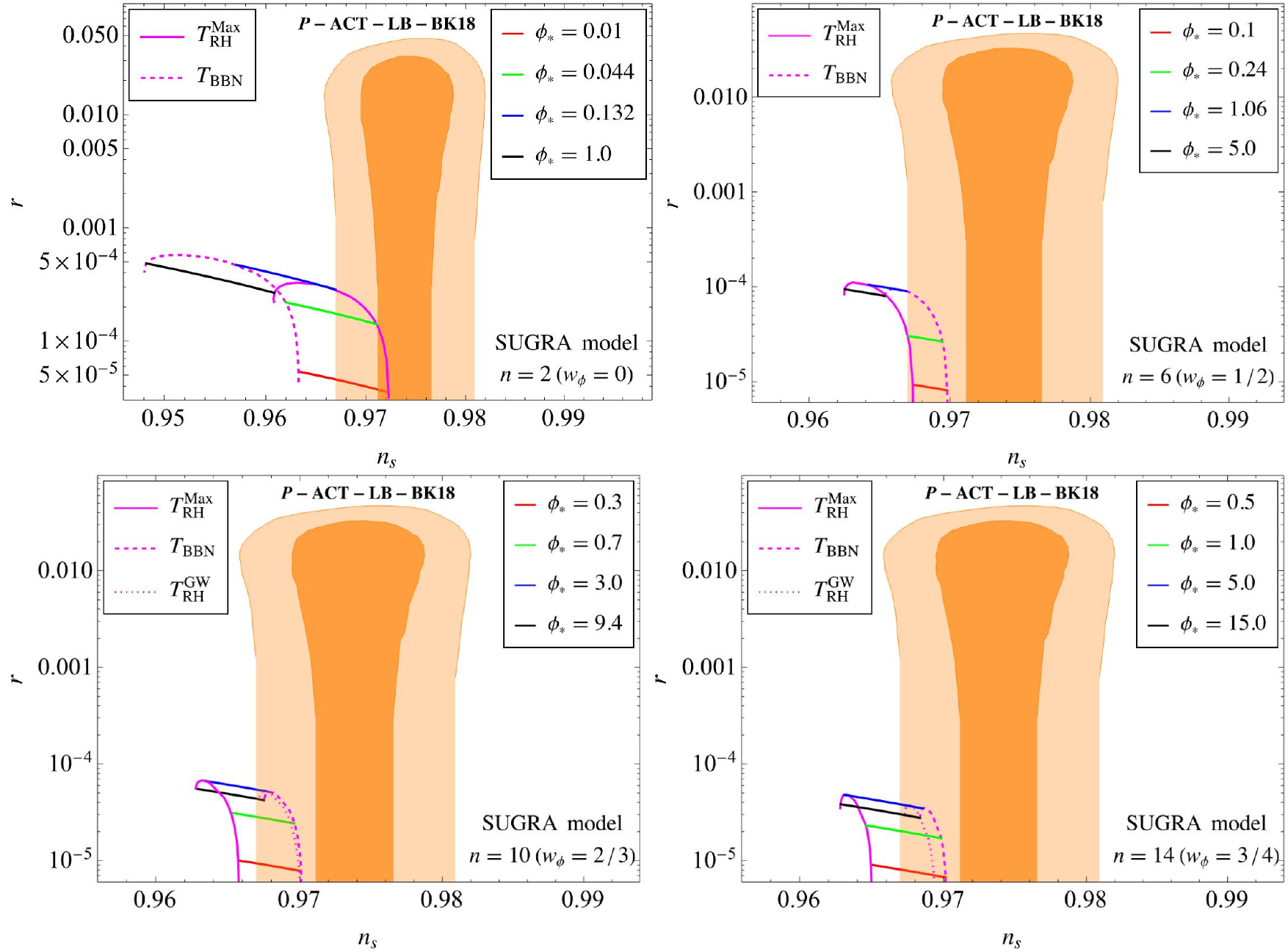}
\caption[]{\justifying \em  Predictions from the SUGRA model for various combinations of $(\phi_*,\, w_\phi)$ are shown in the $(n_s,\, r)$ plane. All other details are identical to those in Fig.~\ref{ns-r:minimal}.}
\label{ns-r:SUGRA}
\end{figure*}

\begin{figure}[!ht]
    \centering
    \includegraphics[scale=0.4]{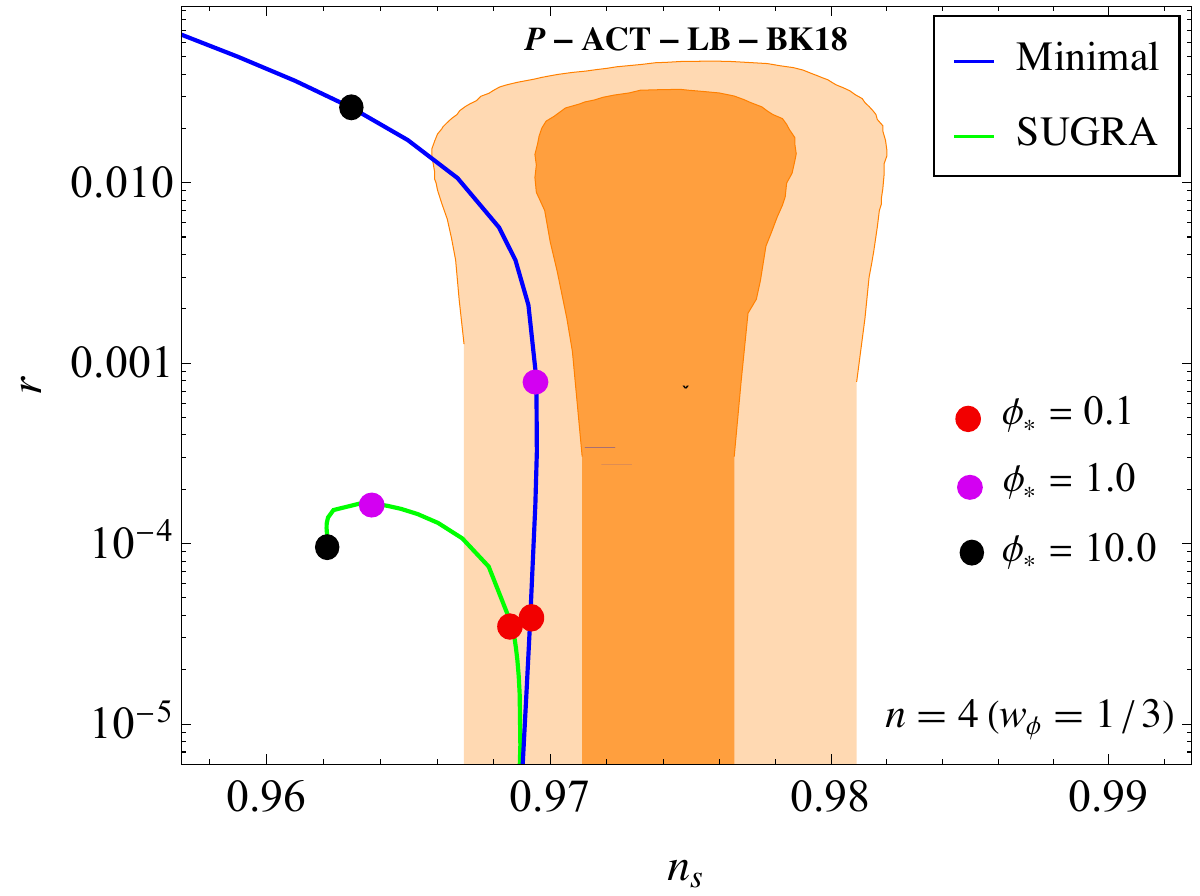}
    \caption[]{\justifying \it The $(n_s,\, r)$ plane shows the predictions of both the Minimal and SUGRA models for $n = 4$, overlaid with the most recent combined constraints from \textsc{P-ACT-LB-BK18}. The shaded regions in dark and light orange represent the $1\sigma$ (68\% confidence level) and $2\sigma$ (95\% confidence level) contours, respectively.}
    \label{fig:w13}
\end{figure}

In the second model, we consider a power-law plateau potential inspired by supergravity. The corresponding inflaton potential, referred to as the SUGRA model \cite{Maity:2019ltu}, can be written as:
\begin{equation}
V_{\rm SUGRA} (\phi) =  \frac{\Lambda \phi^n}{\mbox{Exp}[{-\frac{\phi^2}{2 M_p^2}}]+\left(\frac{\phi}{\phi_*}\right)^n} . 
\end{equation}
Note that in the limit $\phi> M_p >\phi_*$, the above potential boils down to the first one in Eq.\ref{model1}. Therefore, considering $\phi_* \leq {\cal{O}}(1)$ in unit of ${\rm M_p}$, we get an unified expression of the scalar spectral index $(n_s)$, and the tensor scalar ratio $(r)$ in terms of $n$,\,$N_k$, and $\phi_*$, as 
 \bea \label{nsrvsN}
 1 - n_s = 
  \frac{\gamma}{N_k}~~;~~
  r = 
  8n^2 \left(\frac{\phi_*}{\rm M_p}\right)^{\frac{n\gamma}{(n+1)}} [n(n+2)N_k]^{-\gamma}. 
\eea
Where $\gamma = {2(n+1)}/{(n+2)}$. Where inflationary e-folding numner can also be approximately expressed as,
 \begin{align}
       N_k\simeq \frac{\phi_*^2}{n {\rm M_p}^2} \frac{1}{(n+2)} \left(\frac{\phi_k}{\phi_*}\right)^{(n+2)}.
       \label{efold3}
     \end{align}
Here the suffix $k$ represents the associated quantity measured for a particular scale $k$, which we set to be the CMB pivot scale $k_* = 0.05 ~{\rm Mpc}^{-1}$ throughout. From the analytic expressions, it is interesting to observe that with increasing $n$, $n_s$ value decreases. For example, assuming $N_k=50$, $\phi_* =0.01 M_{\rm p}$, one obtains $n_s= (0.969, 0.966, 0.965)$, and $r =(4\times 10^{-5}, 2 \times 10^{-6}, 3 \times 10^{-7})$ for $n = (2,4,6)$ accordingly. From this we can clearly observe that $n=2$ is favored in light of ACT results, and our subsequent detailed analysis indeed concludes the same as illustrated in Figures \ref{ns-r:minimal}, \ref{ns-r:SUGRA}, \ref{fig:w13}, and \ref{fig:Tre_constraints}. 

Beyond driving inflation, the inflaton also decays into radiation during the post-inflationary reheating phase, thereby initiating the hot Big Bang. In the following, we turn to another important constraint arising from the overproduction of primordial gravitational waves (PGWs). These gravitational waves originate from vacuum tensor fluctuations during inflation and evolve through the post-inflationary universe. Their contribution to the effective number of relativistic species, $\Delta N_{\rm eff}$, can be significant for stiff post-inflationary phases ($w_\phi > 1/3$), placing tight constraints on the reheating history and model viability.

\section{Reheating Dynamics and Constraints from Primordial Gravitational Waves}
\label{sec:reheating_GW}
\begin{figure*}[t]  
\includegraphics[width=0017.50cm,height=015cm]{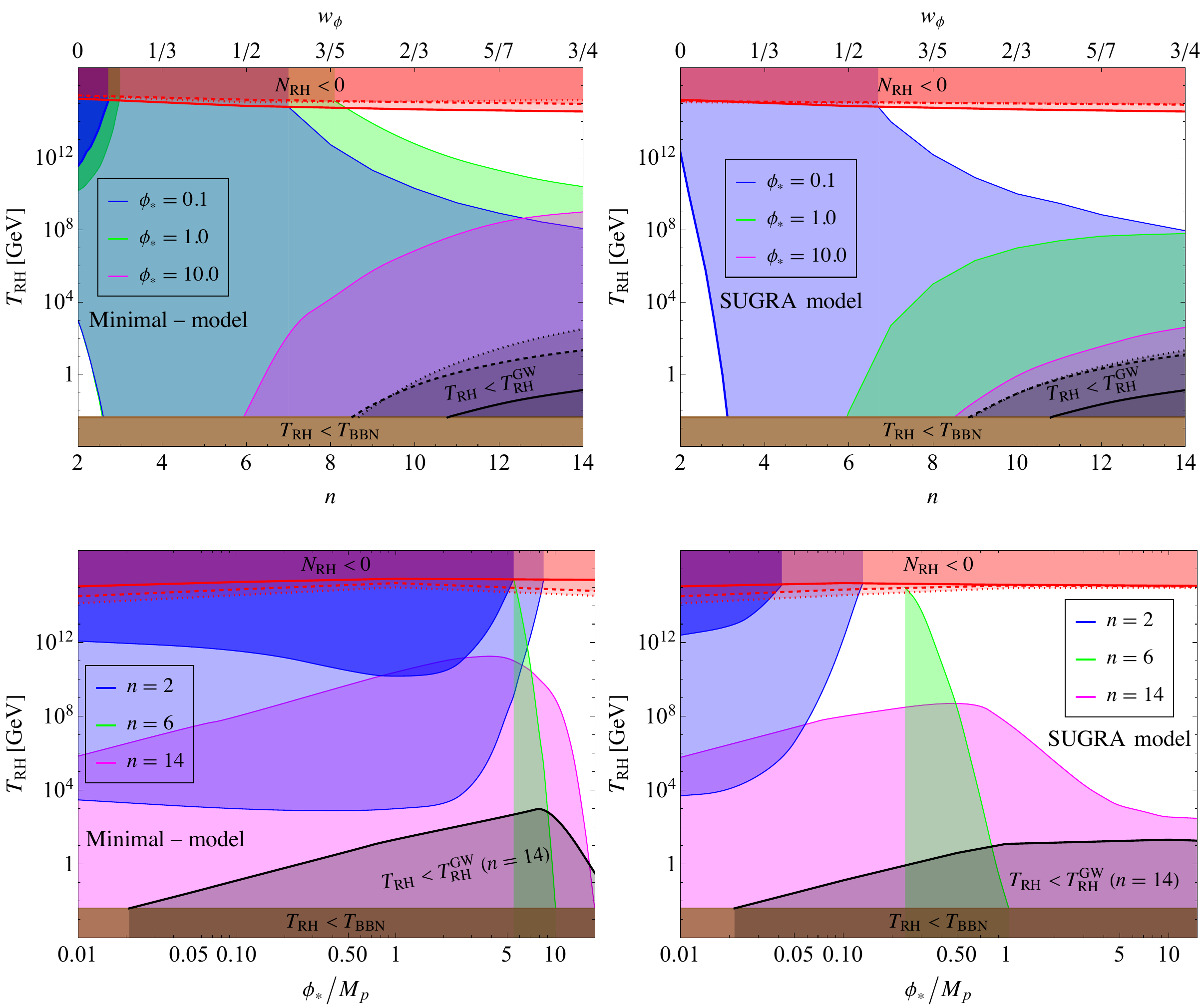}
   \caption[]{\justifying \em \textbf{Upper panel:} Constraints on the reheating temperature based on the $1\sigma$ (68\% C.L.) and $2\sigma$ (95\% C.L.) limits from the recent {\rm P-ACT-LB-BK18} dataset for both the minimal and SUGRA models. The red band indicates the region where $N_{\rm RH} < 0$ for three different values of $\phi_{\ast} = (0.1, 1, 10)$, represented by solid, dashed, and dotted lines, respectively. The brown band highlights the region where $T_{\rm RH} < T_{\rm BBN}$.
\textbf{Lower panel:} Reheating temperature as a function of $\phi_{\ast}$ for different values of $n$, shown for both the minimal and SUGRA models. Note that the dark shaded region represents the parameter space consistent with the $1\sigma$ observational bounds, while the light shaded region corresponds to the $2\sigma$ allowed range. The field value $\phi_{\ast}$ is expressed in units of the reduced Planck mass, $M_{\rm p}$.}
\label{fig:Tre_constraints}
\end{figure*}
\begin{table*}[!ht]
    \centering
    \renewcommand{\arraystretch}{1.2}
    \begin{tabular}{c|c|c|c|c|c|c|c|c}
    \hline
    \hline
        \multirow{2}{*}{$n(\w)$} & \multicolumn{2}{|c}{upper limit on $\phi_{\ast}/M_{\rm p}$} & \multicolumn{2}{|c}{range of $N_{\rm k}$} & \multicolumn{2}{|c}{maximum $T_{\rm RH}$ [GeV]} & \multicolumn{2}{|c}{minimum $T_{\rm RH}$ [GeV]}\\
        \cline{2-9}
        & Minimal & SUGRA & Minimal & SUGRA & Minimal & SUGRA & Minimal & SUGRA\\
        \hline
        $2(0)$ & $8.5$ & $0.132$ & $[45.8-56.5]$ & $[45.6-54.2]$ & $2.7\times10^{15}$ & $1.7\times 10^{15}$ & $8.0\times10^{2}$ & $5\times 10^3$\\
        $6(1/2)$ & $10.1$ & $1.04$ & $[53.0-61.4]$ & $[53.0-59.0]$ & $2.7\times10^{15}$ & $9.7\times10^{14}$ & ${\rm T}_{\rm BBN}$ & ${\rm T}_{\rm BBN}$\\
        $14(3/4)$ & $17.8$ & $20.0$ & $[56.7-61.6]$ & $[56.7-61.6]$ & $1.8\times10^{11}$ & $5\times10^{8}$ & ${\rm T}_{\rm BBN}$ & ${\rm T}_{\rm BBN}$\\
    \hline
    \hline
    \end{tabular}
    \caption[]{\justifying \it Limiting values of the inflationary model parameters ($\phi_{\ast}, N_k$) and corresponding bounds on the reheating temperature, derived using the $2\sigma$ constraints from the recent {\rm P-ACT-LB-BK18} dataset, are shown for both Minimal and SUGRA models.}
    \label{tab:model_constraint}
\end{table*}
\begin{table*}[!ht]
    \centering
    \renewcommand{\arraystretch}{1.2}
    \begin{tabular}{c|c|c|c|c|c|c|c|c}
    \hline
    \hline
        \multirow{2}{*}{$n(\w)$} & \multicolumn{2}{|c}{upper limit on $\phi_{\ast}/M_{\rm p}$} & \multicolumn{2}{|c}{range of $N_{\rm k}$} & \multicolumn{2}{|c}{maximum $T_{\rm RH}$ [GeV]} & \multicolumn{2}{|c}{minimum $T_{\rm RH}$ [GeV]}\\
        \cline{2-9}
        & Minimal & SUGRA & Minimal & SUGRA & Minimal & SUGRA & Minimal & SUGRA\\
        \hline
        $2(0)$ & $5.5$ & $0.042$ & $[55.6-56.5]$ & $[52.2-54.2]$ & $2.7\times10^{15}$ & $1.5\times 10^{15}$ & $1.5\times10^{10}$ & $2.5\times 10^{12}$\\
    \hline
    \hline
    \end{tabular}
    \caption[]{\justifying \it Limiting values of the inflationary model parameters ($\phi_{\ast}, N_k$) and corresponding bounds on the reheating temperature, derived using the $1\sigma$ constraints from the recent {\rm P-ACT-LB-BK18} dataset for $n=2$ (matter-like reheating), are shown for both Minimal and SUGRA models.}
    \label{tab:model_constraint-1sigma}
\end{table*}
As just stated the inflationary epoch is followed by a reheating phase~\cite{Kolb:1990vq,Shtanov:1994ce,Kofman:1997yn,Allahverdi:2010xz,Lozanov:2019jxc,Garcia:2020wiy,Haque:2023yra,Haque:2023zhb} resulting into hot radiation dominated universe (RD). During the entire period of this phase the classical inflaton field coherently oscillates around it minimum with approximated potential $V \propto \phi^n$, and quantum mechanically produces massless radiation fields. Under the generic condition, 
such oscillating field, behaving as an ideal fluid with energy density $\rho_\phi$, is characterized by an effectively constant reheating equation of state (EoS)~\cite{Ford:1986sy, Garcia:2020wiy} as,
\begin{equation}
\w \simeq \frac{n - 2}{n + 2}.
\end{equation} 
Along with this EoS $\w$, the reheating phase is, therefore parametrized by another important parameter associated with the produced radiation energy density called reheating temperature $T_{\rm RH}$ defined at the conclusion of reheating. The reheating concludes when the both the energy density becomes comparable, i.e., $\rho_\phi \simeq \rho_{\rm RH}$. Where, $\rho_{\rm RH}$ is the radiation energy density at the end of reheating. 
Assuming energy density at the end of inflation being $\rho_\phi =\rho_{\rm end} =3 M_{\rm p}^2 H_{\rm end}^2$ one immediately obtains the following relation 
\begin{equation} \label{back}
\frac{\rho_{\rm end}}{\rho_{\rm RH}} = \left( \frac{a_{\rm end}}{a_{\rm RH}} \right)^{-3(1 + \w)} = e^{3(1+\w)N_{\rm RH}},
\end{equation}
where $a_{\rm end}$ and $a_{\rm RH}$ denote the scale factors at the end of inflation and reheating, respectively, and $N_{\rm RH}$ is e-folding number during reheating. In thermal equilibrium the radiation energy density is calculated by utilizing Stefan-Boltzmann relation $\rho_{\rm RH} = \frac{\pi^2}{30} g_{\rm \ast RH}T_{\rm RH}^4$. Replacing this into Eq. \ref{back}, one obtains the expression for the reheating temperature as
\begin{equation}
\label{tre1}
T_{\rm RH} \simeq \left(\frac{90 \Mpl^2 H_{\rm end}^2}{\pi^2 g_{\rm \ast RH}}\right)^{1/4} e^{-\frac{3}{4}N_{\rm RH}(1+\w)}.
\end{equation}
Post reheating period until today is assumed to conserve entropy which results into an important constraint relation between $T_{\rm RH}$ and the present CMB temperature $T_0 (= 2.735 , \mathrm{K})$) as follows ~\cite{Dai:2014jja,Cook:2015vqa},
\begin{equation} \label{tre2}
T_{\rm RH} = \left(\frac{43}{11 g_{*S,\rm RH}}\right)^{1/3} T_0 \frac{H_{\rm k}}{k_{\ast}} e^{-(N_{k} + N_{\rm RH})}.
\end{equation}
Equating equations \ref{tre1}, and \ref{tre2}, one can therefore estimate the inflationary e-folding number through following model-dependent relation between $N_k$ and $T_{\rm RH}$:
\begin{align}
\label{eq:Ninf_reheating}
    N_{k} = \,&\log \left[
\left(\frac{43}{11g_{\ast S, RH}}\right)^{1/3} T_0 \frac{H_{\rm k}}{k_{\ast}}  \right . \nonumber \\
& \times \left. T_{\rm RH}^{\frac{4}{3(1 + \w)} - 1} 
\left(\frac{\pi^2g_{\ast RH}}{90 \Mpl^2 H_{\rm end}^2}\right)^{\frac{1}{3(1 + \w)}}
\right]
\end{align}
where $(g_{\ast S, \rm RH},, g_{\ast \rm RH})$ denote the effective number of relativistic degrees of freedom associated with the entropy and thermal bath, respectively, evaluated at the end of reheating. Given the model parameters $(n, \phi_*)$, and $n_s = 0.9743 \pm 0.0034$ and $r < 0.038$ yield, would fix the inflationary e-folding number along with a fixed reheating temperature depicted in Fig. \ref{fig:Tre_constraints}. 

Another important observable with great potential to constrain the inflation model is PGWs. It is indeed  believed to be a smoking gun signal for an inflationary paradigm. Whereas the strength of the PGW at large scale directly measures the inflationary energy scale through the measurement of $r$, small scale modes, on the other hand, encode the additional scales of the post-inflationary phase. Inflation followed by reheating breaks the scale-invariant property of inflationary PGW spectrum, particularly for high-frequency modes $k>k_{\rm RH}$ as $\Omega^{(0)}_{\rm GW}h^2 \propto k^{2(3\w-1)/(3\w+1)}$ \cite{Haque:2021dha}. Here $k_{\rm RH}$ is the scale entering the horizon right at the end of reheating. Such breaking yields the
spectrum to be red tilted for $\w < 1/3$, and blue tilted for $\w >1/3$. It is this blue tilted spectrum that turns PGWs behaving as an additional relativistic degrees of freedom parameterized by $\Delta N_{\rm eff}$ that can have adverse effect on the Big Bang Nucleosynthesis process. Joint constraints from ACT and Planck therefore place a strict upper bound on $\Delta N_{\rm eff} \leq 0.17$ at 95\% confidence level \cite{ACT:2025fju,ACT:2025tim}, and such bound indeed can further constraint the inflation model under consideration. 

This observational bound translates into the following integral constraint on the PGW energy density:
\begin{align}
\int_{k_{\rm RH}}^{k_{\rm end}}\frac{\d k}{k}\,\Omega^{\rm (0)}_{\rm GW}(k)\,h^2
\leq \frac{7}{8}\,\left(\frac{4}{11}\right)^{4/3}\,\Omega^{\rm (0)}_{\gamma}\,h^2\,\Delta N_\mathrm{eff},
\label{eq:deltaneff}
\end{align}
where $\Omega^{\rm (0)}_{\gamma}\,h^2 \simeq 2.47 \times 10^{-5}$ denotes the present-day photon energy density, and $(k_{\rm RH},\,k_{\rm end})$ represent the comoving wave numbers corresponding to horizon re-entry at the end of reheating and at the end of inflation, respectively. This PGW induced bound yields a constraints on the possible lower bound on the reheating temperature particularly for stiff equation of sate as follows,
\begin{eqnarray}
T_{\rm RH} \gtrsim & \left[\frac{\Omega^{\rm (0)}_{\gamma}\, h^2}{5.61\times 10^{-6}\,\Delta N_{\rm eff}}\,
\frac{H_{\rm end}^2}{12\, \pi^2\, {M_{\rm P}}^2}\, \frac{(1+3\,w_{\phi})}{2\,\pi\,
(3\,w_{\phi} -1)}\,\right]^{\frac{3\,(1+w_{\phi})}{4\,(3\,w_{\phi} -1)}}\, \nonumber \\
 & \times \left(\frac{90\,H_{\rm end}^2\,{M_{\rm p}}^2}{\pi^2\,g_{*\mathrm{RH}}}\right)^{\frac{1}{4}} \equiv T_{\rm RH}^{\rm GW}.
\label{eq:BBNrestriction}
\end{eqnarray}
(for detailed expression, see Refs.~\cite{Maity:2024cpq,Haque:2021dha,Chakraborty:2023ocr}). 
This bound plays a crucial role in constraining specific inflationary models, as we shall examine in the following section.
\section{Results}
 \label{sec:results}
We present a class of plateau inflation models in light of the latest {\rm P-ACT-LB-BK18} dataset, incorporating both inflationary and post-inflationary reheating dynamics. The model is characterized by the potential index $n$, the mass scale $\phi_{\ast}$, and a derived inflationary e-folding number $N_k$, all of which are tightly constrained by the recent bounds on the scalar spectral index $n_s$ and the tensor-to-scalar ratio $r$ (see, for instance Fig. \ref{ns-r:minimal}).
For each value of $n$, we determine the allowed ranges of $(\phi_{\ast},\,N_k)$ and the reheating temperature $T_{\rm RH}$ consistent with the $2\sigma$ observational bounds (see, for instance, Table-\ref{tab:model_constraint} and lower panel of Fig.\ref{fig:Tre_constraints}). We have also performed a complementary analysis by fixing $\phi_{\ast}$ and varying $n$ (equivalently, the post-inflationary equation-of-state parameter $w_\phi$) to examine how the allowed ranges shift under changes in the reheating dynamics (see, for instance, Table-\ref{tab:model_constraint} and upper panel of Fig.\ref{fig:Tre_constraints}). Our results indicate that the minimal model with $n = 2$ ($w_\phi = 0$) is the most favored scenario, admitting a wide window of reheating temperature between $10^3$ GeV and $\mathcal{O}(10^{15})$ GeV, while accommodating $N_k \sim (45.8–56.5)$ falling within $2\sigma$ confidence region in $n_s-r$ plane provided by ACT. As $n$ increases, the allowed parameter space progressively narrows down. Further for stiffer reheating EoS, additional constraint from BBN comes into play, and Consequently become less viable. At the $1\sigma$ level, we find that only models with EoS values close to $w_\phi = 0$ remain compatible with current observational data (for detailed constraints, see Table~\ref{tab:model_constraint-1sigma}).

In parallel, we analyze the supergravity-inspired (SUGRA) extension of the model, which introduces an exponential suppression at large field values. While its inflationary predictions closely resemble those of the minimal case for $\phi_{\ast} < M_{\rm p}$, we find that the SUGRA model parameters are tightly constrained on both $\phi_{\ast}$ and $T_{\rm RH}$ across all $n$, especially when $\phi_{\ast} > M_{\rm p}$ (see, for instance, Figs \ref{ns-r:SUGRA}, and \ref{fig:w13}). Nevertheless, for $n = 2$, both minimal and SUGRA models remain consistent with current observations at both the $2\sigma$ and $1\sigma$ levels.\\
Tables~\ref{tab:model_constraint} and I\ref{tab:model_constraint-1sigma} summarize the viable ranges of $(\phi_{\ast}, N_k)$ and $T_{\rm RH}$ for various values of $n$. Our analysis demonstrates that both the minimal and SUGRA frameworks support reheating scenarios consistent with theoretical expectations and current observations, particularly for models with matter-like or near matter-like post-inflationary dynamics. These results underscore the utility of combining precision CMB data with reheating physics to test inflationary paradigms robustly.
\section{Conclusions}
\label{sec:conclusion}

In this work, we have revisited the minimal plateau inflationary model in light of the recent {\rm P-ACT-LB-BK18} dataset, incorporating constraints from both inflationary observables and post-inflationary reheating dynamics, together with the $\Delta N_{\rm eff}$ bound arising from the overproduction of PGWs. Our analysis explores the viability of the model across a range of potential indices $n$, field scales $\phi_{\ast}$, and reheating equations of state $w_\phi$.
Our main findings are summarized below:
\begin{itemize}
\item The minimal model with $n = 2$ (i.e., $w_\phi = 0$) remains consistent with current observations both in $1\sigma$ and $2\sigma$ confidence regions. At $2\sigma$ C.L., yielding the allowed range for the reheating temperature $T_{\rm RH} \in (10^3, 10^{15}) \,\mbox{GeV}$ and the number of e-folds $N_k \in (46, 56)$.
\item At the $1\sigma$ level, only matter-like reheating scenarios ($w_\phi \approx 0$) or close to that remain compatible with the latest CMB constraints, significantly narrowing the viable parameter space for steeper post-inflationary dynamics.
\item For larger values of $n$, the allowed parameter space becomes increasingly constrained due to the PGW-induced $\Delta N_{\rm eff}$ bound, which imposes a lower limit on $T_{\rm RH}$, particularly for stiff reheating phases ($w_\phi >> 1/3$).
\item The supergravity-inspired extension of the model generally yields tighter constraints on $\phi_{\ast}$ and $T_{\rm RH}$, particularly when $\phi_{\ast} > M_{\rm p}$, and reproduces results similar to the minimal model for $\phi_{\ast} \ll M_{\rm p}$. While it remains compatible with observations for $n = 2$ at both the $1\sigma$ and $2\sigma$ levels, our analysis indicates that the standard minimal model provides a better overall fit to the latest observational data.
\end{itemize}
Looking forward, upcoming precision measurements from CMB-S4 and space-based gravitational wave observatories will further refine the constraints on early-universe dynamics. Extending this framework to include non-standard reheating scenarios or additional relic production mechanisms, such as dark matter and baryogenesis, offers a compelling avenue for future research.
\black
\section*{Acknowledgements}
MRH acknowledges ISI Kolkata for providing financial support through Research Associateship. DM wishes to thank IITG Astro-Gravity group members.

\appendix


\bibliographystyle{apsrev4-2}
\bibliography{references}

\begin{thebibliography}{59}%
\makeatletter
\providecommand \@ifxundefined [1]{%
 \@ifx{#1\undefined}
}%
\providecommand \@ifnum [1]{%
 \ifnum #1\expandafter \@firstoftwo
 \else \expandafter \@secondoftwo
 \fi
}%
\providecommand \@ifx [1]{%
 \ifx #1\expandafter \@firstoftwo
 \else \expandafter \@secondoftwo
 \fi
}%
\providecommand \natexlab [1]{#1}%
\providecommand \enquote  [1]{``#1''}%
\providecommand \bibnamefont  [1]{#1}%
\providecommand \bibfnamefont [1]{#1}%
\providecommand \citenamefont [1]{#1}%
\providecommand \href@noop [0]{\@secondoftwo}%
\providecommand \href [0]{\begingroup \@sanitize@url \@href}%
\providecommand \@href[1]{\@@startlink{#1}\@@href}%
\providecommand \@@href[1]{\endgroup#1\@@endlink}%
\providecommand \@sanitize@url [0]{\catcode `\\12\catcode `\$12\catcode `\&12\catcode `\#12\catcode `\^12\catcode `\_12\catcode `\%12\relax}%
\providecommand \@@startlink[1]{}%
\providecommand \@@endlink[0]{}%
\providecommand \url  [0]{\begingroup\@sanitize@url \@url }%
\providecommand \@url [1]{\endgroup\@href {#1}{\urlprefix }}%
\providecommand \urlprefix  [0]{URL }%
\providecommand \Eprint [0]{\href }%
\providecommand \doibase [0]{https://doi.org/}%
\providecommand \selectlanguage [0]{\@gobble}%
\providecommand \bibinfo  [0]{\@secondoftwo}%
\providecommand \bibfield  [0]{\@secondoftwo}%
\providecommand \translation [1]{[#1]}%
\providecommand \BibitemOpen [0]{}%
\providecommand \bibitemStop [0]{}%
\providecommand \bibitemNoStop [0]{.\EOS\space}%
\providecommand \EOS [0]{\spacefactor3000\relax}%
\providecommand \BibitemShut  [1]{\csname bibitem#1\endcsname}%
\let\auto@bib@innerbib\@empty
\bibitem [{\citenamefont {Louis}\ \emph {et~al.}(2025)\citenamefont {Louis} \emph {et~al.}}]{ACT:2025fju}%
  \BibitemOpen
  \bibfield  {author} {\bibinfo {author} {\bibfnamefont {T.}~\bibnamefont {Louis}} \emph {et~al.} (\bibinfo {collaboration} {ACT}),\ }\href@noop {} {\  (\bibinfo {year} {2025})},\ \Eprint {https://arxiv.org/abs/2503.14452} {arXiv:2503.14452 [astro-ph.CO]} \BibitemShut {NoStop}%
\bibitem [{\citenamefont {Calabrese}\ \emph {et~al.}(2025)\citenamefont {Calabrese} \emph {et~al.}}]{ACT:2025tim}%
  \BibitemOpen
  \bibfield  {author} {\bibinfo {author} {\bibfnamefont {E.}~\bibnamefont {Calabrese}} \emph {et~al.} (\bibinfo {collaboration} {ACT}),\ }\href@noop {} {\  (\bibinfo {year} {2025})},\ \Eprint {https://arxiv.org/abs/2503.14454} {arXiv:2503.14454 [astro-ph.CO]} \BibitemShut {NoStop}%
\bibitem [{\citenamefont {Abdul~Karim}\ \emph {et~al.}(2025)\citenamefont {Abdul~Karim} \emph {et~al.}}]{DESI:2025zgx}%
  \BibitemOpen
  \bibfield  {author} {\bibinfo {author} {\bibfnamefont {M.}~\bibnamefont {Abdul~Karim}} \emph {et~al.} (\bibinfo {collaboration} {DESI}),\ }\href@noop {} {\  (\bibinfo {year} {2025})},\ \Eprint {https://arxiv.org/abs/2503.14738} {arXiv:2503.14738 [astro-ph.CO]} \BibitemShut {NoStop}%
\bibitem [{\citenamefont {Linde}(1983)}]{Linde:1983gd}%
  \BibitemOpen
  \bibfield  {author} {\bibinfo {author} {\bibfnamefont {A.~D.}\ \bibnamefont {Linde}},\ }\href {https://doi.org/10.1016/0370-2693(83)90837-7} {\bibfield  {journal} {\bibinfo  {journal} {Phys. Lett. B}\ }\textbf {\bibinfo {volume} {129}},\ \bibinfo {pages} {177} (\bibinfo {year} {1983})}\BibitemShut {NoStop}%
\bibitem [{\citenamefont {Martin}\ \emph {et~al.}(2014)\citenamefont {Martin}, \citenamefont {Ringeval},\ and\ \citenamefont {Vennin}}]{Martin:2013tda}%
  \BibitemOpen
  \bibfield  {author} {\bibinfo {author} {\bibfnamefont {J.}~\bibnamefont {Martin}}, \bibinfo {author} {\bibfnamefont {C.}~\bibnamefont {Ringeval}},\ and\ \bibinfo {author} {\bibfnamefont {V.}~\bibnamefont {Vennin}},\ }\href {https://doi.org/10.1016/j.dark.2024.101653} {\bibfield  {journal} {\bibinfo  {journal} {Phys. Dark Univ.}\ }\textbf {\bibinfo {volume} {5-6}},\ \bibinfo {pages} {75} (\bibinfo {year} {2014})},\ \Eprint {https://arxiv.org/abs/1303.3787} {arXiv:1303.3787 [astro-ph.CO]} \BibitemShut {NoStop}%
\bibitem [{\citenamefont {Linde}(1994)}]{Linde:1993cn}%
  \BibitemOpen
  \bibfield  {author} {\bibinfo {author} {\bibfnamefont {A.~D.}\ \bibnamefont {Linde}},\ }\href {https://doi.org/10.1103/PhysRevD.49.748} {\bibfield  {journal} {\bibinfo  {journal} {Phys. Rev. D}\ }\textbf {\bibinfo {volume} {49}},\ \bibinfo {pages} {748} (\bibinfo {year} {1994})},\ \Eprint {https://arxiv.org/abs/astro-ph/9307002} {arXiv:astro-ph/9307002} \BibitemShut {NoStop}%
\bibitem [{\citenamefont {Starobinsky}(1980)}]{Starobinsky:1980te}%
  \BibitemOpen
  \bibfield  {author} {\bibinfo {author} {\bibfnamefont {A.~A.}\ \bibnamefont {Starobinsky}},\ }\href {https://doi.org/10.1016/0370-2693(80)90670-X} {\bibfield  {journal} {\bibinfo  {journal} {Phys. Lett. B}\ }\textbf {\bibinfo {volume} {91}},\ \bibinfo {pages} {99} (\bibinfo {year} {1980})}\BibitemShut {NoStop}%
\bibitem [{\citenamefont {Kallosh}\ and\ \citenamefont {Linde}(2013)}]{Kallosh:2013hoa}%
  \BibitemOpen
  \bibfield  {author} {\bibinfo {author} {\bibfnamefont {R.}~\bibnamefont {Kallosh}}\ and\ \bibinfo {author} {\bibfnamefont {A.}~\bibnamefont {Linde}},\ }\href {https://doi.org/10.1088/1475-7516/2013/07/002} {\bibfield  {journal} {\bibinfo  {journal} {JCAP}\ }\textbf {\bibinfo {volume} {07}},\ \bibinfo {pages} {002}},\ \Eprint {https://arxiv.org/abs/1306.5220} {arXiv:1306.5220 [hep-th]} \BibitemShut {NoStop}%
\bibitem [{\citenamefont {Ellis}\ \emph {et~al.}(2013)\citenamefont {Ellis}, \citenamefont {Nanopoulos},\ and\ \citenamefont {Olive}}]{Ellis:2013nxa}%
  \BibitemOpen
  \bibfield  {author} {\bibinfo {author} {\bibfnamefont {J.}~\bibnamefont {Ellis}}, \bibinfo {author} {\bibfnamefont {D.~V.}\ \bibnamefont {Nanopoulos}},\ and\ \bibinfo {author} {\bibfnamefont {K.~A.}\ \bibnamefont {Olive}},\ }\href {https://doi.org/10.1088/1475-7516/2013/10/009} {\bibfield  {journal} {\bibinfo  {journal} {JCAP}\ }\textbf {\bibinfo {volume} {10}},\ \bibinfo {pages} {009}},\ \Eprint {https://arxiv.org/abs/1307.3537} {arXiv:1307.3537 [hep-th]} \BibitemShut {NoStop}%
\bibitem [{\citenamefont {Kallosh}\ \emph {et~al.}(2013)\citenamefont {Kallosh}, \citenamefont {Linde},\ and\ \citenamefont {Roest}}]{Kallosh:2013yoa}%
  \BibitemOpen
  \bibfield  {author} {\bibinfo {author} {\bibfnamefont {R.}~\bibnamefont {Kallosh}}, \bibinfo {author} {\bibfnamefont {A.}~\bibnamefont {Linde}},\ and\ \bibinfo {author} {\bibfnamefont {D.}~\bibnamefont {Roest}},\ }\href {https://doi.org/10.1007/JHEP11(2013)198} {\bibfield  {journal} {\bibinfo  {journal} {JHEP}\ }\textbf {\bibinfo {volume} {11}},\ \bibinfo {pages} {198}},\ \Eprint {https://arxiv.org/abs/1311.0472} {arXiv:1311.0472 [hep-th]} \BibitemShut {NoStop}%
\bibitem [{\citenamefont {Freese}\ \emph {et~al.}(1990)\citenamefont {Freese}, \citenamefont {Frieman},\ and\ \citenamefont {Olinto}}]{Freese:1990rb}%
  \BibitemOpen
  \bibfield  {author} {\bibinfo {author} {\bibfnamefont {K.}~\bibnamefont {Freese}}, \bibinfo {author} {\bibfnamefont {J.~A.}\ \bibnamefont {Frieman}},\ and\ \bibinfo {author} {\bibfnamefont {A.~V.}\ \bibnamefont {Olinto}},\ }\href {https://doi.org/10.1103/PhysRevLett.65.3233} {\bibfield  {journal} {\bibinfo  {journal} {Phys. Rev. Lett.}\ }\textbf {\bibinfo {volume} {65}},\ \bibinfo {pages} {3233} (\bibinfo {year} {1990})}\BibitemShut {NoStop}%
\bibitem [{\citenamefont {Roest}(2014)}]{Roest:2013fha}%
  \BibitemOpen
  \bibfield  {author} {\bibinfo {author} {\bibfnamefont {D.}~\bibnamefont {Roest}},\ }\href {https://doi.org/10.1088/1475-7516/2014/01/007} {\bibfield  {journal} {\bibinfo  {journal} {JCAP}\ }\textbf {\bibinfo {volume} {01}},\ \bibinfo {pages} {007}},\ \Eprint {https://arxiv.org/abs/1309.1285} {arXiv:1309.1285 [hep-th]} \BibitemShut {NoStop}%
\bibitem [{\citenamefont {Ferrara}\ \emph {et~al.}(2013)\citenamefont {Ferrara}, \citenamefont {Kallosh}, \citenamefont {Linde},\ and\ \citenamefont {Porrati}}]{Ferrara:2013rsa}%
  \BibitemOpen
  \bibfield  {author} {\bibinfo {author} {\bibfnamefont {S.}~\bibnamefont {Ferrara}}, \bibinfo {author} {\bibfnamefont {R.}~\bibnamefont {Kallosh}}, \bibinfo {author} {\bibfnamefont {A.}~\bibnamefont {Linde}},\ and\ \bibinfo {author} {\bibfnamefont {M.}~\bibnamefont {Porrati}},\ }\href {https://doi.org/10.1103/PhysRevD.88.085038} {\bibfield  {journal} {\bibinfo  {journal} {Phys. Rev. D}\ }\textbf {\bibinfo {volume} {88}},\ \bibinfo {pages} {085038} (\bibinfo {year} {2013})},\ \Eprint {https://arxiv.org/abs/1307.7696} {arXiv:1307.7696 [hep-th]} \BibitemShut {NoStop}%
\bibitem [{\citenamefont {Kallosh}\ \emph {et~al.}(2014{\natexlab{a}})\citenamefont {Kallosh}, \citenamefont {Linde},\ and\ \citenamefont {Roest}}]{Kallosh:2013tua}%
  \BibitemOpen
  \bibfield  {author} {\bibinfo {author} {\bibfnamefont {R.}~\bibnamefont {Kallosh}}, \bibinfo {author} {\bibfnamefont {A.}~\bibnamefont {Linde}},\ and\ \bibinfo {author} {\bibfnamefont {D.}~\bibnamefont {Roest}},\ }\href {https://doi.org/10.1103/PhysRevLett.112.011303} {\bibfield  {journal} {\bibinfo  {journal} {Phys. Rev. Lett.}\ }\textbf {\bibinfo {volume} {112}},\ \bibinfo {pages} {011303} (\bibinfo {year} {2014}{\natexlab{a}})},\ \Eprint {https://arxiv.org/abs/1310.3950} {arXiv:1310.3950 [hep-th]} \BibitemShut {NoStop}%
\bibitem [{\citenamefont {Cecotti}\ and\ \citenamefont {Kallosh}(2014)}]{Cecotti:2014ipa}%
  \BibitemOpen
  \bibfield  {author} {\bibinfo {author} {\bibfnamefont {S.}~\bibnamefont {Cecotti}}\ and\ \bibinfo {author} {\bibfnamefont {R.}~\bibnamefont {Kallosh}},\ }\href {https://doi.org/10.1007/JHEP05(2014)114} {\bibfield  {journal} {\bibinfo  {journal} {JHEP}\ }\textbf {\bibinfo {volume} {05}},\ \bibinfo {pages} {114}},\ \Eprint {https://arxiv.org/abs/1403.2932} {arXiv:1403.2932 [hep-th]} \BibitemShut {NoStop}%
\bibitem [{\citenamefont {Kallosh}\ \emph {et~al.}(2014{\natexlab{b}})\citenamefont {Kallosh}, \citenamefont {Linde},\ and\ \citenamefont {Roest}}]{Kallosh:2014rga}%
  \BibitemOpen
  \bibfield  {author} {\bibinfo {author} {\bibfnamefont {R.}~\bibnamefont {Kallosh}}, \bibinfo {author} {\bibfnamefont {A.}~\bibnamefont {Linde}},\ and\ \bibinfo {author} {\bibfnamefont {D.}~\bibnamefont {Roest}},\ }\href {https://doi.org/10.1007/JHEP08(2014)052} {\bibfield  {journal} {\bibinfo  {journal} {JHEP}\ }\textbf {\bibinfo {volume} {08}},\ \bibinfo {pages} {052}},\ \Eprint {https://arxiv.org/abs/1405.3646} {arXiv:1405.3646 [hep-th]} \BibitemShut {NoStop}%
\bibitem [{\citenamefont {Galante}\ \emph {et~al.}(2015)\citenamefont {Galante}, \citenamefont {Kallosh}, \citenamefont {Linde},\ and\ \citenamefont {Roest}}]{Galante:2014ifa}%
  \BibitemOpen
  \bibfield  {author} {\bibinfo {author} {\bibfnamefont {M.}~\bibnamefont {Galante}}, \bibinfo {author} {\bibfnamefont {R.}~\bibnamefont {Kallosh}}, \bibinfo {author} {\bibfnamefont {A.}~\bibnamefont {Linde}},\ and\ \bibinfo {author} {\bibfnamefont {D.}~\bibnamefont {Roest}},\ }\href {https://doi.org/10.1103/PhysRevLett.114.141302} {\bibfield  {journal} {\bibinfo  {journal} {Phys. Rev. Lett.}\ }\textbf {\bibinfo {volume} {114}},\ \bibinfo {pages} {141302} (\bibinfo {year} {2015})},\ \Eprint {https://arxiv.org/abs/1412.3797} {arXiv:1412.3797 [hep-th]} \BibitemShut {NoStop}%
\bibitem [{\citenamefont {Bennett}\ \emph {et~al.}(1996)\citenamefont {Bennett}, \citenamefont {Banday}, \citenamefont {Gorski}, \citenamefont {Hinshaw}, \citenamefont {Jackson}, \citenamefont {Keegstra}, \citenamefont {Kogut}, \citenamefont {Smoot}, \citenamefont {Wilkinson},\ and\ \citenamefont {Wright}}]{Bennett:1996ce}%
  \BibitemOpen
  \bibfield  {author} {\bibinfo {author} {\bibfnamefont {C.~L.}\ \bibnamefont {Bennett}}, \bibinfo {author} {\bibfnamefont {A.}~\bibnamefont {Banday}}, \bibinfo {author} {\bibfnamefont {K.~M.}\ \bibnamefont {Gorski}}, \bibinfo {author} {\bibfnamefont {G.}~\bibnamefont {Hinshaw}}, \bibinfo {author} {\bibfnamefont {P.}~\bibnamefont {Jackson}}, \bibinfo {author} {\bibfnamefont {P.}~\bibnamefont {Keegstra}}, \bibinfo {author} {\bibfnamefont {A.}~\bibnamefont {Kogut}}, \bibinfo {author} {\bibfnamefont {G.~F.}\ \bibnamefont {Smoot}}, \bibinfo {author} {\bibfnamefont {D.~T.}\ \bibnamefont {Wilkinson}},\ and\ \bibinfo {author} {\bibfnamefont {E.~L.}\ \bibnamefont {Wright}},\ }\href {https://doi.org/10.1086/310075} {\bibfield  {journal} {\bibinfo  {journal} {Astrophys. J. Lett.}\ }\textbf {\bibinfo {volume} {464}},\ \bibinfo {pages} {L1} (\bibinfo {year} {1996})},\ \Eprint {https://arxiv.org/abs/astro-ph/9601067} {arXiv:astro-ph/9601067} \BibitemShut {NoStop}%
\bibitem [{\citenamefont {Hauser}\ \emph {et~al.}(1998)\citenamefont {Hauser} \emph {et~al.}}]{Hauser:1998ri}%
  \BibitemOpen
  \bibfield  {author} {\bibinfo {author} {\bibfnamefont {M.~G.}\ \bibnamefont {Hauser}} \emph {et~al.},\ }\href {https://doi.org/10.1086/306379} {\bibfield  {journal} {\bibinfo  {journal} {Astrophys. J.}\ }\textbf {\bibinfo {volume} {508}},\ \bibinfo {pages} {25} (\bibinfo {year} {1998})},\ \Eprint {https://arxiv.org/abs/astro-ph/9806167} {arXiv:astro-ph/9806167} \BibitemShut {NoStop}%
\bibitem [{\citenamefont {Kallosh}\ \emph {et~al.}(2025)\citenamefont {Kallosh}, \citenamefont {Linde},\ and\ \citenamefont {Roest}}]{Kallosh:2025rni}%
  \BibitemOpen
  \bibfield  {author} {\bibinfo {author} {\bibfnamefont {R.}~\bibnamefont {Kallosh}}, \bibinfo {author} {\bibfnamefont {A.}~\bibnamefont {Linde}},\ and\ \bibinfo {author} {\bibfnamefont {D.}~\bibnamefont {Roest}},\ }\href@noop {} {\  (\bibinfo {year} {2025})},\ \Eprint {https://arxiv.org/abs/2503.21030} {arXiv:2503.21030 [hep-th]} \BibitemShut {NoStop}%
\bibitem [{\citenamefont {Aoki}\ \emph {et~al.}(2025)\citenamefont {Aoki}, \citenamefont {Otsuka},\ and\ \citenamefont {Yanagita}}]{Aoki:2025wld}%
  \BibitemOpen
  \bibfield  {author} {\bibinfo {author} {\bibfnamefont {S.}~\bibnamefont {Aoki}}, \bibinfo {author} {\bibfnamefont {H.}~\bibnamefont {Otsuka}},\ and\ \bibinfo {author} {\bibfnamefont {R.}~\bibnamefont {Yanagita}},\ }\href@noop {} {\  (\bibinfo {year} {2025})},\ \Eprint {https://arxiv.org/abs/2504.01622} {arXiv:2504.01622 [hep-ph]} \BibitemShut {NoStop}%
\bibitem [{\citenamefont {Berera}\ \emph {et~al.}(2025)\citenamefont {Berera}, \citenamefont {Brahma}, \citenamefont {Qiu}, \citenamefont {O.~Ramos},\ and\ \citenamefont {Rodrigues}}]{Berera:2025vsu}%
  \BibitemOpen
  \bibfield  {author} {\bibinfo {author} {\bibfnamefont {A.}~\bibnamefont {Berera}}, \bibinfo {author} {\bibfnamefont {S.}~\bibnamefont {Brahma}}, \bibinfo {author} {\bibfnamefont {Z.}~\bibnamefont {Qiu}}, \bibinfo {author} {\bibfnamefont {R.}~\bibnamefont {O.~Ramos}},\ and\ \bibinfo {author} {\bibfnamefont {G.~S.}\ \bibnamefont {Rodrigues}},\ }\href@noop {} {\  (\bibinfo {year} {2025})},\ \Eprint {https://arxiv.org/abs/2504.02655} {arXiv:2504.02655 [hep-th]} \BibitemShut {NoStop}%
\bibitem [{\citenamefont {Dioguardi}\ \emph {et~al.}(2025)\citenamefont {Dioguardi}, \citenamefont {Iovino},\ and\ \citenamefont {Racioppi}}]{Dioguardi:2025vci}%
  \BibitemOpen
  \bibfield  {author} {\bibinfo {author} {\bibfnamefont {C.}~\bibnamefont {Dioguardi}}, \bibinfo {author} {\bibfnamefont {A.~J.}\ \bibnamefont {Iovino}},\ and\ \bibinfo {author} {\bibfnamefont {A.}~\bibnamefont {Racioppi}},\ }\href@noop {} {\  (\bibinfo {year} {2025})},\ \Eprint {https://arxiv.org/abs/2504.02809} {arXiv:2504.02809 [gr-qc]} \BibitemShut {NoStop}%
\bibitem [{\citenamefont {Gialamas}\ \emph {et~al.}(2025{\natexlab{a}})\citenamefont {Gialamas}, \citenamefont {Karam}, \citenamefont {Racioppi},\ and\ \citenamefont {Raidal}}]{Gialamas:2025kef}%
  \BibitemOpen
  \bibfield  {author} {\bibinfo {author} {\bibfnamefont {I.~D.}\ \bibnamefont {Gialamas}}, \bibinfo {author} {\bibfnamefont {A.}~\bibnamefont {Karam}}, \bibinfo {author} {\bibfnamefont {A.}~\bibnamefont {Racioppi}},\ and\ \bibinfo {author} {\bibfnamefont {M.}~\bibnamefont {Raidal}},\ }\href@noop {} {\  (\bibinfo {year} {2025}{\natexlab{a}})},\ \Eprint {https://arxiv.org/abs/2504.06002} {arXiv:2504.06002 [astro-ph.CO]} \BibitemShut {NoStop}%
\bibitem [{\citenamefont {Salvio}(2025)}]{Salvio:2025izr}%
  \BibitemOpen
  \bibfield  {author} {\bibinfo {author} {\bibfnamefont {A.}~\bibnamefont {Salvio}},\ }\href@noop {} {\  (\bibinfo {year} {2025})},\ \Eprint {https://arxiv.org/abs/2504.10488} {arXiv:2504.10488 [hep-ph]} \BibitemShut {NoStop}%
\bibitem [{\citenamefont {Antoniadis}\ \emph {et~al.}(2025)\citenamefont {Antoniadis}, \citenamefont {Ellis}, \citenamefont {Ke}, \citenamefont {Nanopoulos},\ and\ \citenamefont {Olive}}]{Antoniadis:2025pfa}%
  \BibitemOpen
  \bibfield  {author} {\bibinfo {author} {\bibfnamefont {I.}~\bibnamefont {Antoniadis}}, \bibinfo {author} {\bibfnamefont {J.}~\bibnamefont {Ellis}}, \bibinfo {author} {\bibfnamefont {W.}~\bibnamefont {Ke}}, \bibinfo {author} {\bibfnamefont {D.~V.}\ \bibnamefont {Nanopoulos}},\ and\ \bibinfo {author} {\bibfnamefont {K.~A.}\ \bibnamefont {Olive}},\ }\href@noop {} {\  (\bibinfo {year} {2025})},\ \Eprint {https://arxiv.org/abs/2504.12283} {arXiv:2504.12283 [hep-ph]} \BibitemShut {NoStop}%
\bibitem [{\citenamefont {Kim}\ \emph {et~al.}(2025)\citenamefont {Kim}, \citenamefont {Wang}, \citenamefont {Zhang},\ and\ \citenamefont {Ren}}]{Kim:2025dyi}%
  \BibitemOpen
  \bibfield  {author} {\bibinfo {author} {\bibfnamefont {J.}~\bibnamefont {Kim}}, \bibinfo {author} {\bibfnamefont {X.}~\bibnamefont {Wang}}, \bibinfo {author} {\bibfnamefont {Y.-l.}\ \bibnamefont {Zhang}},\ and\ \bibinfo {author} {\bibfnamefont {Z.}~\bibnamefont {Ren}},\ }\href@noop {} {\  (\bibinfo {year} {2025})},\ \Eprint {https://arxiv.org/abs/2504.12035} {arXiv:2504.12035 [astro-ph.CO]} \BibitemShut {NoStop}%
\bibitem [{\citenamefont {Dioguardi}\ and\ \citenamefont {Karam}(2025)}]{Dioguardi:2025mpp}%
  \BibitemOpen
  \bibfield  {author} {\bibinfo {author} {\bibfnamefont {C.}~\bibnamefont {Dioguardi}}\ and\ \bibinfo {author} {\bibfnamefont {A.}~\bibnamefont {Karam}},\ }\href@noop {} {\  (\bibinfo {year} {2025})},\ \Eprint {https://arxiv.org/abs/2504.12937} {arXiv:2504.12937 [gr-qc]} \BibitemShut {NoStop}%
\bibitem [{\citenamefont {Gao}\ \emph {et~al.}(2025)\citenamefont {Gao}, \citenamefont {Gong}, \citenamefont {Yi},\ and\ \citenamefont {Zhang}}]{Gao:2025onc}%
  \BibitemOpen
  \bibfield  {author} {\bibinfo {author} {\bibfnamefont {Q.}~\bibnamefont {Gao}}, \bibinfo {author} {\bibfnamefont {Y.}~\bibnamefont {Gong}}, \bibinfo {author} {\bibfnamefont {Z.}~\bibnamefont {Yi}},\ and\ \bibinfo {author} {\bibfnamefont {F.}~\bibnamefont {Zhang}},\ }\href@noop {} {\  (\bibinfo {year} {2025})},\ \Eprint {https://arxiv.org/abs/2504.15218} {arXiv:2504.15218 [astro-ph.CO]} \BibitemShut {NoStop}%
\bibitem [{\citenamefont {He}\ \emph {et~al.}(2025)\citenamefont {He}, \citenamefont {Hong},\ and\ \citenamefont {Mukaida}}]{He:2025bli}%
  \BibitemOpen
  \bibfield  {author} {\bibinfo {author} {\bibfnamefont {M.}~\bibnamefont {He}}, \bibinfo {author} {\bibfnamefont {M.}~\bibnamefont {Hong}},\ and\ \bibinfo {author} {\bibfnamefont {K.}~\bibnamefont {Mukaida}},\ }\href@noop {} {\  (\bibinfo {year} {2025})},\ \Eprint {https://arxiv.org/abs/2504.16069} {arXiv:2504.16069 [astro-ph.CO]} \BibitemShut {NoStop}%
\bibitem [{\citenamefont {Pallis}(2025)}]{Pallis:2025epn}%
  \BibitemOpen
  \bibfield  {author} {\bibinfo {author} {\bibfnamefont {C.}~\bibnamefont {Pallis}}\ }(\bibinfo {year} {2025})\ \Eprint {https://arxiv.org/abs/2504.20273} {arXiv:2504.20273 [hep-ph]} \BibitemShut {NoStop}%
\bibitem [{\citenamefont {Drees}\ and\ \citenamefont {Xu}(2025)}]{Drees:2025ngb}%
  \BibitemOpen
  \bibfield  {author} {\bibinfo {author} {\bibfnamefont {M.}~\bibnamefont {Drees}}\ and\ \bibinfo {author} {\bibfnamefont {Y.}~\bibnamefont {Xu}},\ }\href@noop {} {\  (\bibinfo {year} {2025})},\ \Eprint {https://arxiv.org/abs/2504.20757} {arXiv:2504.20757 [astro-ph.CO]} \BibitemShut {NoStop}%
\bibitem [{\citenamefont {Haque}\ \emph {et~al.}(2025{\natexlab{a}})\citenamefont {Haque}, \citenamefont {Pal},\ and\ \citenamefont {Paul}}]{Haque:2025uis}%
  \BibitemOpen
  \bibfield  {author} {\bibinfo {author} {\bibfnamefont {M.~R.}\ \bibnamefont {Haque}}, \bibinfo {author} {\bibfnamefont {S.}~\bibnamefont {Pal}},\ and\ \bibinfo {author} {\bibfnamefont {D.}~\bibnamefont {Paul}},\ }\href@noop {} {\  (\bibinfo {year} {2025}{\natexlab{a}})},\ \Eprint {https://arxiv.org/abs/2505.04615} {arXiv:2505.04615 [astro-ph.CO]} \BibitemShut {NoStop}%
\bibitem [{\citenamefont {Haque}\ \emph {et~al.}(2025{\natexlab{b}})\citenamefont {Haque}, \citenamefont {Pal},\ and\ \citenamefont {Paul}}]{Haque:2025uri}%
  \BibitemOpen
  \bibfield  {author} {\bibinfo {author} {\bibfnamefont {M.~R.}\ \bibnamefont {Haque}}, \bibinfo {author} {\bibfnamefont {S.}~\bibnamefont {Pal}},\ and\ \bibinfo {author} {\bibfnamefont {D.}~\bibnamefont {Paul}},\ }\href@noop {} {\  (\bibinfo {year} {2025}{\natexlab{b}})},\ \Eprint {https://arxiv.org/abs/2505.01517} {arXiv:2505.01517 [astro-ph.CO]} \BibitemShut {NoStop}%
\bibitem [{\citenamefont {Yin}(2025)}]{Yin:2025rrs}%
  \BibitemOpen
  \bibfield  {author} {\bibinfo {author} {\bibfnamefont {W.}~\bibnamefont {Yin}},\ }\href@noop {} {\  (\bibinfo {year} {2025})},\ \Eprint {https://arxiv.org/abs/2505.03004} {arXiv:2505.03004 [hep-ph]} \BibitemShut {NoStop}%
\bibitem [{\citenamefont {Byrnes}\ \emph {et~al.}(2025)\citenamefont {Byrnes}, \citenamefont {Cort\^es},\ and\ \citenamefont {Liddle}}]{Byrnes:2025kit}%
  \BibitemOpen
  \bibfield  {author} {\bibinfo {author} {\bibfnamefont {C.~T.}\ \bibnamefont {Byrnes}}, \bibinfo {author} {\bibfnamefont {M.}~\bibnamefont {Cort\^es}},\ and\ \bibinfo {author} {\bibfnamefont {A.~R.}\ \bibnamefont {Liddle}},\ }\href@noop {} {\  (\bibinfo {year} {2025})},\ \Eprint {https://arxiv.org/abs/2505.09682} {arXiv:2505.09682 [astro-ph.CO]} \BibitemShut {NoStop}%
\bibitem [{\citenamefont {Maity}(2025)}]{Maity:2025czp}%
  \BibitemOpen
  \bibfield  {author} {\bibinfo {author} {\bibfnamefont {S.}~\bibnamefont {Maity}},\ }\href@noop {} {\  (\bibinfo {year} {2025})},\ \Eprint {https://arxiv.org/abs/2505.10534} {arXiv:2505.10534 [astro-ph.CO]} \BibitemShut {NoStop}%
\bibitem [{\citenamefont {Mondal}\ \emph {et~al.}(2025)\citenamefont {Mondal}, \citenamefont {Mondal},\ and\ \citenamefont {Chakraborty}}]{Mondal:2025kur}%
  \BibitemOpen
  \bibfield  {author} {\bibinfo {author} {\bibfnamefont {R.}~\bibnamefont {Mondal}}, \bibinfo {author} {\bibfnamefont {S.}~\bibnamefont {Mondal}},\ and\ \bibinfo {author} {\bibfnamefont {A.}~\bibnamefont {Chakraborty}},\ }\href@noop {} {\  (\bibinfo {year} {2025})},\ \Eprint {https://arxiv.org/abs/2505.13387} {arXiv:2505.13387 [hep-ph]} \BibitemShut {NoStop}%
\bibitem [{\citenamefont {Peng}\ \emph {et~al.}(2025)\citenamefont {Peng}, \citenamefont {Chen},\ and\ \citenamefont {Liu}}]{Peng:2025bws}%
  \BibitemOpen
  \bibfield  {author} {\bibinfo {author} {\bibfnamefont {Z.-Z.}\ \bibnamefont {Peng}}, \bibinfo {author} {\bibfnamefont {Z.-C.}\ \bibnamefont {Chen}},\ and\ \bibinfo {author} {\bibfnamefont {L.}~\bibnamefont {Liu}},\ }\href@noop {} {\  (\bibinfo {year} {2025})},\ \Eprint {https://arxiv.org/abs/2505.12816} {arXiv:2505.12816 [astro-ph.CO]} \BibitemShut {NoStop}%
\bibitem [{\citenamefont {Yi}\ \emph {et~al.}(2025)\citenamefont {Yi}, \citenamefont {Wang}, \citenamefont {Gao},\ and\ \citenamefont {Gong}}]{Yi:2025dms}%
  \BibitemOpen
  \bibfield  {author} {\bibinfo {author} {\bibfnamefont {Z.}~\bibnamefont {Yi}}, \bibinfo {author} {\bibfnamefont {X.}~\bibnamefont {Wang}}, \bibinfo {author} {\bibfnamefont {Q.}~\bibnamefont {Gao}},\ and\ \bibinfo {author} {\bibfnamefont {Y.}~\bibnamefont {Gong}},\ }\href@noop {} {\  (\bibinfo {year} {2025})},\ \Eprint {https://arxiv.org/abs/2505.10268} {arXiv:2505.10268 [astro-ph.CO]} \BibitemShut {NoStop}%
\bibitem [{\citenamefont {Gialamas}\ \emph {et~al.}(2025{\natexlab{b}})\citenamefont {Gialamas}, \citenamefont {Katsoulas},\ and\ \citenamefont {Tamvakis}}]{Gialamas:2025ofz}%
  \BibitemOpen
  \bibfield  {author} {\bibinfo {author} {\bibfnamefont {I.~D.}\ \bibnamefont {Gialamas}}, \bibinfo {author} {\bibfnamefont {T.}~\bibnamefont {Katsoulas}},\ and\ \bibinfo {author} {\bibfnamefont {K.}~\bibnamefont {Tamvakis}},\ }\href@noop {} {\  (\bibinfo {year} {2025}{\natexlab{b}})},\ \Eprint {https://arxiv.org/abs/2505.03608} {arXiv:2505.03608 [gr-qc]} \BibitemShut {NoStop}%
\bibitem [{\citenamefont {Yogesh}\ \emph {et~al.}(2025)\citenamefont {Yogesh}, \citenamefont {Mohammadi}, \citenamefont {Wu},\ and\ \citenamefont {Zhu}}]{Yogesh:2025wak}%
  \BibitemOpen
  \bibfield  {author} {\bibinfo {author} {\bibnamefont {Yogesh}}, \bibinfo {author} {\bibfnamefont {A.}~\bibnamefont {Mohammadi}}, \bibinfo {author} {\bibfnamefont {Q.}~\bibnamefont {Wu}},\ and\ \bibinfo {author} {\bibfnamefont {T.}~\bibnamefont {Zhu}},\ }\href@noop {} {\  (\bibinfo {year} {2025})},\ \Eprint {https://arxiv.org/abs/2505.05363} {arXiv:2505.05363 [astro-ph.CO]} \BibitemShut {NoStop}%
\bibitem [{\citenamefont {Kallosh}\ and\ \citenamefont {Linde}(2025)}]{Kallosh:2025ijd}%
  \BibitemOpen
  \bibfield  {author} {\bibinfo {author} {\bibfnamefont {R.}~\bibnamefont {Kallosh}}\ and\ \bibinfo {author} {\bibfnamefont {A.}~\bibnamefont {Linde}},\ }\href@noop {} {\  (\bibinfo {year} {2025})},\ \Eprint {https://arxiv.org/abs/2505.13646} {arXiv:2505.13646 [hep-th]} \BibitemShut {NoStop}%
\bibitem [{\citenamefont {Maity}\ and\ \citenamefont {Saha}(2019)}]{Maity:2019ltu}%
  \BibitemOpen
  \bibfield  {author} {\bibinfo {author} {\bibfnamefont {D.}~\bibnamefont {Maity}}\ and\ \bibinfo {author} {\bibfnamefont {P.}~\bibnamefont {Saha}},\ }\href {https://doi.org/10.1088/1361-6382/ab0038} {\bibfield  {journal} {\bibinfo  {journal} {Class. Quant. Grav.}\ }\textbf {\bibinfo {volume} {36}},\ \bibinfo {pages} {045010} (\bibinfo {year} {2019})},\ \Eprint {https://arxiv.org/abs/1902.01895} {arXiv:1902.01895 [gr-qc]} \BibitemShut {NoStop}%
\bibitem [{\citenamefont {Maity}(2017)}]{Maity:2016zeu}%
  \BibitemOpen
  \bibfield  {author} {\bibinfo {author} {\bibfnamefont {D.}~\bibnamefont {Maity}},\ }\href {https://doi.org/10.1016/j.nuclphysb.2017.04.005} {\bibfield  {journal} {\bibinfo  {journal} {Nucl. Phys. B}\ }\textbf {\bibinfo {volume} {919}},\ \bibinfo {pages} {560} (\bibinfo {year} {2017})},\ \Eprint {https://arxiv.org/abs/1606.08179} {arXiv:1606.08179 [hep-ph]} \BibitemShut {NoStop}%
\bibitem [{\citenamefont {Kolb}\ and\ \citenamefont {Turner}(2019)}]{Kolb:1990vq}%
  \BibitemOpen
  \bibfield  {author} {\bibinfo {author} {\bibfnamefont {E.~W.}\ \bibnamefont {Kolb}}\ and\ \bibinfo {author} {\bibfnamefont {M.~S.}\ \bibnamefont {Turner}},\ }\href {https://doi.org/10.1201/9780429492860} {\emph {\bibinfo {title} {{The Early Universe}}}},\ Vol.~\bibinfo {volume} {69}\ (\bibinfo  {publisher} {Taylor and Francis},\ \bibinfo {year} {2019})\BibitemShut {NoStop}%
\bibitem [{\citenamefont {Shtanov}\ \emph {et~al.}(1995)\citenamefont {Shtanov}, \citenamefont {Traschen},\ and\ \citenamefont {Brandenberger}}]{Shtanov:1994ce}%
  \BibitemOpen
  \bibfield  {author} {\bibinfo {author} {\bibfnamefont {Y.}~\bibnamefont {Shtanov}}, \bibinfo {author} {\bibfnamefont {J.~H.}\ \bibnamefont {Traschen}},\ and\ \bibinfo {author} {\bibfnamefont {R.~H.}\ \bibnamefont {Brandenberger}},\ }\href {https://doi.org/10.1103/PhysRevD.51.5438} {\bibfield  {journal} {\bibinfo  {journal} {Phys. Rev. D}\ }\textbf {\bibinfo {volume} {51}},\ \bibinfo {pages} {5438} (\bibinfo {year} {1995})},\ \Eprint {https://arxiv.org/abs/hep-ph/9407247} {arXiv:hep-ph/9407247} \BibitemShut {NoStop}%
\bibitem [{\citenamefont {Kofman}\ \emph {et~al.}(1997)\citenamefont {Kofman}, \citenamefont {Linde},\ and\ \citenamefont {Starobinsky}}]{Kofman:1997yn}%
  \BibitemOpen
  \bibfield  {author} {\bibinfo {author} {\bibfnamefont {L.}~\bibnamefont {Kofman}}, \bibinfo {author} {\bibfnamefont {A.~D.}\ \bibnamefont {Linde}},\ and\ \bibinfo {author} {\bibfnamefont {A.~A.}\ \bibnamefont {Starobinsky}},\ }\href {https://doi.org/10.1103/PhysRevD.56.3258} {\bibfield  {journal} {\bibinfo  {journal} {Phys. Rev. D}\ }\textbf {\bibinfo {volume} {56}},\ \bibinfo {pages} {3258} (\bibinfo {year} {1997})},\ \Eprint {https://arxiv.org/abs/hep-ph/9704452} {arXiv:hep-ph/9704452} \BibitemShut {NoStop}%
\bibitem [{\citenamefont {Allahverdi}\ \emph {et~al.}(2010)\citenamefont {Allahverdi}, \citenamefont {Brandenberger}, \citenamefont {Cyr-Racine},\ and\ \citenamefont {Mazumdar}}]{Allahverdi:2010xz}%
  \BibitemOpen
  \bibfield  {author} {\bibinfo {author} {\bibfnamefont {R.}~\bibnamefont {Allahverdi}}, \bibinfo {author} {\bibfnamefont {R.}~\bibnamefont {Brandenberger}}, \bibinfo {author} {\bibfnamefont {F.-Y.}\ \bibnamefont {Cyr-Racine}},\ and\ \bibinfo {author} {\bibfnamefont {A.}~\bibnamefont {Mazumdar}},\ }\href {https://doi.org/10.1146/annurev.nucl.012809.104511} {\bibfield  {journal} {\bibinfo  {journal} {Ann. Rev. Nucl. Part. Sci.}\ }\textbf {\bibinfo {volume} {60}},\ \bibinfo {pages} {27} (\bibinfo {year} {2010})},\ \Eprint {https://arxiv.org/abs/1001.2600} {arXiv:1001.2600 [hep-th]} \BibitemShut {NoStop}%
\bibitem [{\citenamefont {Lozanov}(2019)}]{Lozanov:2019jxc}%
  \BibitemOpen
  \bibfield  {author} {\bibinfo {author} {\bibfnamefont {K.~D.}\ \bibnamefont {Lozanov}},\ }\href@noop {} {\  (\bibinfo {year} {2019})},\ \Eprint {https://arxiv.org/abs/1907.04402} {arXiv:1907.04402 [astro-ph.CO]} \BibitemShut {NoStop}%
\bibitem [{\citenamefont {Garcia}\ \emph {et~al.}(2021)\citenamefont {Garcia}, \citenamefont {Kaneta}, \citenamefont {Mambrini},\ and\ \citenamefont {Olive}}]{Garcia:2020wiy}%
  \BibitemOpen
  \bibfield  {author} {\bibinfo {author} {\bibfnamefont {M.~A.~G.}\ \bibnamefont {Garcia}}, \bibinfo {author} {\bibfnamefont {K.}~\bibnamefont {Kaneta}}, \bibinfo {author} {\bibfnamefont {Y.}~\bibnamefont {Mambrini}},\ and\ \bibinfo {author} {\bibfnamefont {K.~A.}\ \bibnamefont {Olive}},\ }\href {https://doi.org/10.1088/1475-7516/2021/04/012} {\bibfield  {journal} {\bibinfo  {journal} {JCAP}\ }\textbf {\bibinfo {volume} {04}},\ \bibinfo {pages} {012}},\ \Eprint {https://arxiv.org/abs/2012.10756} {arXiv:2012.10756 [hep-ph]} \BibitemShut {NoStop}%
\bibitem [{\citenamefont {Haque}\ \emph {et~al.}(2023)\citenamefont {Haque}, \citenamefont {Maity},\ and\ \citenamefont {Mondal}}]{Haque:2023yra}%
  \BibitemOpen
  \bibfield  {author} {\bibinfo {author} {\bibfnamefont {M.~R.}\ \bibnamefont {Haque}}, \bibinfo {author} {\bibfnamefont {D.}~\bibnamefont {Maity}},\ and\ \bibinfo {author} {\bibfnamefont {R.}~\bibnamefont {Mondal}},\ }\href {https://doi.org/10.1007/JHEP09(2023)012} {\bibfield  {journal} {\bibinfo  {journal} {JHEP}\ }\textbf {\bibinfo {volume} {09}},\ \bibinfo {pages} {012}},\ \Eprint {https://arxiv.org/abs/2301.01641} {arXiv:2301.01641 [hep-ph]} \BibitemShut {NoStop}%
\bibitem [{\citenamefont {Haque}\ \emph {et~al.}(2024)\citenamefont {Haque}, \citenamefont {Maity},\ and\ \citenamefont {Mondal}}]{Haque:2023zhb}%
  \BibitemOpen
  \bibfield  {author} {\bibinfo {author} {\bibfnamefont {M.~R.}\ \bibnamefont {Haque}}, \bibinfo {author} {\bibfnamefont {D.}~\bibnamefont {Maity}},\ and\ \bibinfo {author} {\bibfnamefont {R.}~\bibnamefont {Mondal}},\ }\href {https://doi.org/10.1103/PhysRevD.109.063543} {\bibfield  {journal} {\bibinfo  {journal} {Phys. Rev. D}\ }\textbf {\bibinfo {volume} {109}},\ \bibinfo {pages} {063543} (\bibinfo {year} {2024})},\ \Eprint {https://arxiv.org/abs/2311.07684} {arXiv:2311.07684 [hep-ph]} \BibitemShut {NoStop}%
\bibitem [{\citenamefont {Ford}(1987)}]{Ford:1986sy}%
  \BibitemOpen
  \bibfield  {author} {\bibinfo {author} {\bibfnamefont {L.~H.}\ \bibnamefont {Ford}},\ }\href {https://doi.org/10.1103/PhysRevD.35.2955} {\bibfield  {journal} {\bibinfo  {journal} {Phys. Rev. D}\ }\textbf {\bibinfo {volume} {35}},\ \bibinfo {pages} {2955} (\bibinfo {year} {1987})}\BibitemShut {NoStop}%
\bibitem [{\citenamefont {Dai}\ \emph {et~al.}(2014)\citenamefont {Dai}, \citenamefont {Kamionkowski},\ and\ \citenamefont {Wang}}]{Dai:2014jja}%
  \BibitemOpen
  \bibfield  {author} {\bibinfo {author} {\bibfnamefont {L.}~\bibnamefont {Dai}}, \bibinfo {author} {\bibfnamefont {M.}~\bibnamefont {Kamionkowski}},\ and\ \bibinfo {author} {\bibfnamefont {J.}~\bibnamefont {Wang}},\ }\href {https://doi.org/10.1103/PhysRevLett.113.041302} {\bibfield  {journal} {\bibinfo  {journal} {Phys. Rev. Lett.}\ }\textbf {\bibinfo {volume} {113}},\ \bibinfo {pages} {041302} (\bibinfo {year} {2014})},\ \Eprint {https://arxiv.org/abs/1404.6704} {arXiv:1404.6704 [astro-ph.CO]} \BibitemShut {NoStop}%
\bibitem [{\citenamefont {Cook}\ \emph {et~al.}(2015)\citenamefont {Cook}, \citenamefont {Dimastrogiovanni}, \citenamefont {Easson},\ and\ \citenamefont {Krauss}}]{Cook:2015vqa}%
  \BibitemOpen
  \bibfield  {author} {\bibinfo {author} {\bibfnamefont {J.~L.}\ \bibnamefont {Cook}}, \bibinfo {author} {\bibfnamefont {E.}~\bibnamefont {Dimastrogiovanni}}, \bibinfo {author} {\bibfnamefont {D.~A.}\ \bibnamefont {Easson}},\ and\ \bibinfo {author} {\bibfnamefont {L.~M.}\ \bibnamefont {Krauss}},\ }\href {https://doi.org/10.1088/1475-7516/2015/04/047} {\bibfield  {journal} {\bibinfo  {journal} {JCAP}\ }\textbf {\bibinfo {volume} {04}},\ \bibinfo {pages} {047}},\ \Eprint {https://arxiv.org/abs/1502.04673} {arXiv:1502.04673 [astro-ph.CO]} \BibitemShut {NoStop}%
\bibitem [{\citenamefont {Haque}\ \emph {et~al.}(2021)\citenamefont {Haque}, \citenamefont {Maity}, \citenamefont {Paul},\ and\ \citenamefont {Sriramkumar}}]{Haque:2021dha}%
  \BibitemOpen
  \bibfield  {author} {\bibinfo {author} {\bibfnamefont {M.~R.}\ \bibnamefont {Haque}}, \bibinfo {author} {\bibfnamefont {D.}~\bibnamefont {Maity}}, \bibinfo {author} {\bibfnamefont {T.}~\bibnamefont {Paul}},\ and\ \bibinfo {author} {\bibfnamefont {L.}~\bibnamefont {Sriramkumar}},\ }\href {https://doi.org/10.1103/PhysRevD.104.063513} {\bibfield  {journal} {\bibinfo  {journal} {Phys. Rev. D}\ }\textbf {\bibinfo {volume} {104}},\ \bibinfo {pages} {063513} (\bibinfo {year} {2021})},\ \Eprint {https://arxiv.org/abs/2105.09242} {arXiv:2105.09242 [astro-ph.CO]} \BibitemShut {NoStop}%
\bibitem [{\citenamefont {Maity}\ and\ \citenamefont {Haque}(2024)}]{Maity:2024cpq}%
  \BibitemOpen
  \bibfield  {author} {\bibinfo {author} {\bibfnamefont {S.}~\bibnamefont {Maity}}\ and\ \bibinfo {author} {\bibfnamefont {M.~R.}\ \bibnamefont {Haque}},\ }\href@noop {} {\  (\bibinfo {year} {2024})},\ \Eprint {https://arxiv.org/abs/2407.18246} {arXiv:2407.18246 [astro-ph.CO]} \BibitemShut {NoStop}%
\bibitem [{\citenamefont {Chakraborty}\ \emph {et~al.}(2023)\citenamefont {Chakraborty}, \citenamefont {Haque}, \citenamefont {Maity},\ and\ \citenamefont {Mondal}}]{Chakraborty:2023ocr}%
  \BibitemOpen
  \bibfield  {author} {\bibinfo {author} {\bibfnamefont {A.}~\bibnamefont {Chakraborty}}, \bibinfo {author} {\bibfnamefont {M.~R.}\ \bibnamefont {Haque}}, \bibinfo {author} {\bibfnamefont {D.}~\bibnamefont {Maity}},\ and\ \bibinfo {author} {\bibfnamefont {R.}~\bibnamefont {Mondal}},\ }\href {https://doi.org/10.1103/PhysRevD.108.023515} {\bibfield  {journal} {\bibinfo  {journal} {Phys. Rev. D}\ }\textbf {\bibinfo {volume} {108}},\ \bibinfo {pages} {023515} (\bibinfo {year} {2023})},\ \Eprint {https://arxiv.org/abs/2304.13637} {arXiv:2304.13637 [astro-ph.CO]} \BibitemShut {NoStop}%
\end{thebibliography}%
\end{document}